\documentclass[aps,prb,reprint,showpacs,superscriptaddress]{revtex4-1}

\usepackage{graphicx}% Include figure files
\usepackage{dcolumn}% Align table columns on decimal point
\usepackage{bm}% bold math
\usepackage{amsmath,amssymb}
\usepackage{comment}

\def\U#1{{%
\def\O{\mbox{O}}
\def\u{\mbox{u}}
\mathcode`\u=\mu
\mathcode`\O=\Omega
\mathrm{#1}}}

\def\Re{\mathop{\mathrm{Re}}}
\def\Im{\mathop{\mathrm{Im}}}

\def\nn{\nonumber}

\def\ii{{\mathrm{i}}}
\def\dd{{\mathrm{d}}}
\def\sub#1{_{\scriptsize\mbox{#1}}}

\begin{document}

%\preprint{}

\title{Electromagnetically induced transparency like 
transmission in a metamaterial 
composed of cut-wire pairs with indirect coupling}
% Force line breaks with \\

\author{Yasuhiro Tamayama}
\email{tamayama@vos.nagaokaut.ac.jp}
\author{Kanji Yasui}
\affiliation{Department of Electrical Engineering, Nagaoka University of
Technology, 1603-1 Kamitomioka, Nagaoka, Niigata 940-2188, Japan}
\author{Toshihiro Nakanishi}
\author{Masao Kitano}
 \affiliation{Department of Electronic Science and
 Engineering, Kyoto University, Kyoto 615-8510, Japan}

\date{\today}% It is always \today, today,
             %  but any date may be explicitly specified

\begin{abstract}

We theoretically and numerically investigate 
metamaterials composed of coupled resonators with
indirect coupling. 
First, we theoretically analyze a mechanical model 
of coupled resonators with indirect coupling.  
The theoretical analysis shows that an electromagnetically induced
transparency (EIT)-like phenomenon with a transparency bandwidth narrower
than the resonance linewidths of the constitutive resonators can occur 
in the metamaterial with strong indirect coupling. 
We then numerically 
examine the characteristics of 
the metamaterial composed of coupled cut-wire pairs
using a finite-difference time-domain (FDTD) method. 
The FDTD simulation confirms that 
an EIT-like transparency phenomenon occurs
in the metamaterial owing to indirect coupling.
Finally, we compare the results of the theoretical and
numerical analyses. 
The behavior of the EIT-like metamaterial is found to be
well described by the
mechanical model of the coupled resonators. 

% (Abstract should be about 5% of the length of the article, but less
%  than 500 words)
% about less than 200 words

\end{abstract}

\pacs{78.67.Pt, 41.20.Jb, 78.20.Ci}% PACS, the Physics and Astronomy
                             % Classification Scheme.
%\keywords{Suggested keywords}%Use showkeys class option if keyword
                              %display desired
\maketitle

\section{Introduction}

Metamaterials are arrays of artificial
structures that are much smaller than the wavelength of electromagnetic
waves. The macroscopic characteristics of metamaterials are determined by their
constitutive elements and, therefore, electromagnetic media with desired
properties can be created by designing the shape, material, and density of
the constitutive elements. We can fabricate various media that do not occur
in nature and can use such designed and fabricated metamaterials to control electromagnetic waves at will.

Resonant structures such as split-ring resonators\cite{pendry99} and
electric-field-coupled inductor--capacitor resonators\cite{schurig06_apl}
are typically used as constitutive elements of metamaterials. These structures are
often called meta-atoms. 
The effective relative permittivity and permeability of metamaterials can be varied from
unity and can even be made negative 
by using the strong response of these meta-atoms around the resonant
frequencies. These resonant meta-atoms 
enable the realization of exotic phenomena 
such as negative refraction,\cite{shelby01,valentine08} 
simultaneous negative phase
and group velocities,\cite{dolling06} subwavelength
imaging,\cite{lagarkov04} and cloaking.\cite{schurig06}
In addition, metamaterials composed of coupled resonators, which are
often called metamolecules, have been
investigated to
realize useful characteristics 
such as giant optical activity\cite{liu07_prb}
and enhancement of second-harmonic
generation.\cite{kanazawa11,nakanishi12,czaplicki13} 
Dispersion control by introducing 
coupling between neighboring unit cells has also
been studied.\cite{shamonina08,nakata12}

Although only electric and magnetic direct couplings (near-field
couplings) have been introduced
in the above mentioned metamaterials, indirect coupling mediated by
radiative modes has been
introduced in other coupled-resonator systems. For example, indirect
coupling has been introduced to achieve 
electromagnetically induced transparency (EIT)-like transmission in
a two-dimensional photonic crystal waveguide 
coupled with two resonators,\cite{suh04}
EIT-like scattering and superscattering in a double-slit structure in
a metal film,\cite{verslegers12} and
control of light at the nanoscale using plasmonic
antennas.\cite{zhang_s_12} If indirect coupling could be introduced into
metamaterials as well as these isolated coupled-resonator systems,
further development of methods for controlling electromagnetic waves could
be expected. 
However, it is not obvious whether indirect coupling can be induced only
between meta-atoms in each unit cell of 
metamaterials composed of periodically arranged coupled
resonators (metamolecules). 

In this paper, we analyze the characteristics of
metamaterials composed of coupled resonators with indirect coupling. 
First, we use a mechanical model of coupled resonators to show
that an EIT-like transparency phenomenon occurs when
indirect coupling is introduced. The characteristics of the
EIT-like metamaterial with indirect coupling are compared to those of
previously investigated EIT-like metamaterials with direct
coupling.\cite{zhang_prl08,tassin_prl09,liu_nat09,tamayama10,zhang_apl_10,kurter11,tamayama12}
Then, we show through a finite-difference time-domain (FDTD) 
simulation\cite{taflove05} 
that the EIT-like transparency phenomenon caused by the
indirect coupling is observed in the metamaterial composed of coupled
cut-wire pairs. 
Finally, we discuss whether 
indirect coupling can be induced only between meta-atoms in each unit
cell of metamaterials 
by comparing the theory based on the mechanical model and the results of
the numerical analysis.

\section{Models of coupled resonators}

\begin{figure}[tb]
\begin{center}
\includegraphics[scale=1]{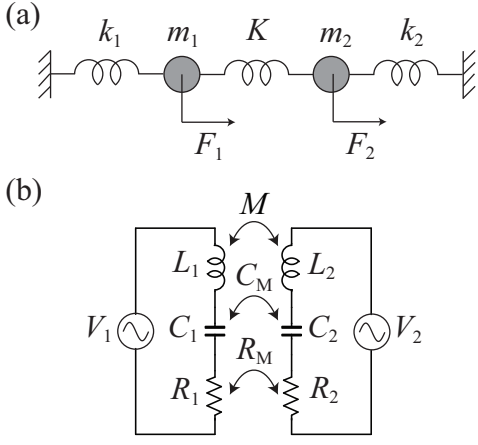}
\caption{(a) Mechanical model and (b) electrical circuit model of
 coupled resonators. }
\label{fig:model}
\end{center}
\end{figure}

We analyze the electromagnetic 
response of a metamaterial composed of metamolecules that are
modeled by the mechanical model shown in Fig.\,\ref{fig:model}(a). 
When the masses of the two particles are the same ($m_1 = m_2 = m$) 
and the elastic constant of the central spring is much smaller than that 
of the other springs ($K \ll k_{1,2}$), the equation of motion is
given by
\begin{align}
&
(-\omega^2 + \omega_1^2 - \ii \gamma_1 \omega) x_1 - \kappa^2 x_2 = \alpha_1
 E_0, \label{eq:10} \\
&
(-\omega^2 + \omega_2^2 - \ii \gamma_2 \omega) x_2 - \kappa^2 x_1 = \alpha_2
 E_0, \label{eq:20}
\end{align}
where $\omega_{1,2} = \sqrt{k_{1,2} / m}$, $\kappa = \sqrt{K/m}$,
$\gamma_{1,2}$ are damping constants, and $F_{1,2} = m \alpha_{1,2} E_0$ 
are external
forces, which are assumed to be proportional to the electric
field $E_0$ of the electromagnetic wave with proportionality constants of
$\alpha_{1,2}$. 
Solving Eqs.\,(\ref{eq:10}) and (\ref{eq:20}) for $x_1$ and $x_2$, 
the following equation is obtained: 
\begin{align}
\begin{bmatrix}
x_1\\
x_2
\end{bmatrix}
&=
\frac{-1}{(\omega^2 - \omega_1^2 + \ii \gamma_1 \omega)(\omega^2 -
\omega_2^2 + \ii \gamma_2 \omega) - \kappa^4} \nn \\
& \hspace{5mm} \times
\begin{bmatrix}
(\omega^2 - \omega_2^2 + \ii \gamma_2 \omega)\alpha_1 - \kappa^2 \alpha_2 \\
(\omega^2 - \omega_1^2 + \ii \gamma_1 \omega)\alpha_2 - \kappa^2 \alpha_1
\end{bmatrix}
E_0 . \label{eq:30}
\end{align}

We assume that 
the electric susceptibility of the metamaterial is proportional
to $x_1 + x_2$. 
We further assume that the characteristics of the two 
resonators are similar to each other and that 
particles 1 and 2 are coupled only via indirect coupling, that is,
$\kappa^2$ is a purely imaginary quantity, $\kappa^2 = \ii
\Im{(\kappa^2)}$. 
When $\gamma_{1,2}$ are derived only from the radiations of the
meta-atoms, both $\gamma_{1,2}$ and $\alpha_{1,2}$ represent the
coupling between the meta-atoms and freespace. Thus,
$\alpha_1/\alpha_2 = \gamma_1 / \gamma_2$ can be safely
assumed.\cite{zhang_s_12} 
Using the above assumptions, we obtain $x_1 + x_2$ at around $\omega
\simeq \omega_0$ as follows: 
\begin{align}
x_1 + x_2 
&\approx 
-2\alpha E_0 ( \omega^2 - \omega_0^2 + \ii \gamma\sub{L} \omega) \nn \\
& \hspace{5mm}
\times \{ (\omega^2 - \omega_0^2 + \ii \gamma_1 \omega)(\omega^2 - \omega_0^2
+ \ii \gamma_2 \omega) \nn \\ 
& \hspace{5mm} 
- (\omega_0^2 - \omega_1^2)(\omega_2^2 -
\omega_0^2) + [\Im{(\kappa^2)}]^2 \}^{-1} 
, \label{eq:40}
\end{align}
where $\omega_0^2=(\alpha_2 \omega_1^2 + \alpha_1
\omega_2^2)/ (\alpha_1 + \alpha_2)$, $ \ii 2 \gamma\sub{L} \omega
\alpha = \ii \gamma_2 \omega \alpha_1 + \ii \gamma_1 \omega \alpha_2 -
(\alpha_1 + \alpha_2) \kappa^2$, and $\alpha = (\alpha_1 + \alpha_2)/2 $.
The right-hand side of Eq.\,(\ref{eq:40}) resembles the susceptibility of
the classical model of EIT with direct
coupling.\cite{alzar02}
The correspondences of main parameters in the susceptibility of  
classical EIT with direct coupling to parameters in
Eq.\,(\ref{eq:40}) are as follows: 
Both the resonant angular 
frequencies of the bright and dark modes are equal to $\omega_0$, the
loss of the dark mode is nearly the same as 
$\gamma\sub{L}$, and the coupling factor between the bright and
dark modes corresponds to $(\omega_0^2 - \omega_1^2)(\omega_2^2 -
\omega_0^2) - [\Im{(\kappa^2)}]^2$. 
This implies that an EIT-like transparency phenomenon occurs at the 
angular frequency $\omega=\omega_0 \approx (\omega_1 + \omega_2)/2$. 
Note that $\gamma\sub{L}$ becomes less than $\gamma_{1,2}$ if
$\Im{(\kappa^2 )} >0$.
From the passive condition for the metamaterial, 
the imaginary part of $x_1 + x_2$ must
be non-negative. 
To satisfy the passive condition at $\omega = \omega_0$ irrespective of
$\omega_1$ and $\omega_2$, $\Im{(\kappa^2)} \leq \omega_0
\sqrt{\gamma_1 \gamma_2}$ is required. 

Next, we consider the group index at $\omega = \omega_0$. 
When the absolute value of $\varDelta = \omega_1 - \omega_2$ is larger
than a certain value $\varDelta\sub{max}$, the transmission bandwidth
decreases; that is, the group index increases with decreasing
$| \varDelta |$. When $| \varDelta |$ is smaller than
$\varDelta\sub{max}$, 
the transmission window gradually disappears and the group index decreases
with decreasing $| \varDelta |$ due to the loss. 
The value of $\varDelta\sub{max}$ can be regarded
as the minimum transmission bandwidth. 
We calculate $\varDelta\sub{max}$ that gives the condition for the largest
group index below. 

We may assume that 
the value of $| \varDelta |$ that maximizes $\Re{[\dd (x_1 + x_2)/\dd \omega |_{\omega =
\omega_0}]}$ is almost equal to $\varDelta\sub{max}$ 
in the case of strongly dispersive media as in the present case. 
For simplicity of the analysis, 
we further assume that $\gamma_1 = \gamma_2 = \gamma_0$, from which
$\Im{(\kappa^2)} = \omega_0 (\gamma_0 - \gamma\sub{L})$ is obtained. 
This implies that 
$\gamma\sub{L}$ that appears in Eq.\,(\ref{eq:40}) 
represents the leak of the indirect
coupling. 
From Eq.\,(\ref{eq:40}), we obtain
\begin{equation}
\Re{\left[ \left. \frac{\dd (x_1 + x_2)}{\dd \omega} \right|_{\omega =
 \omega_0} \right] }
\approx
\frac{4 \alpha E_0(\varDelta^2 - \gamma\sub{L}^2)}{\omega_0(
\varDelta^2 + 2\gamma_0 \gamma\sub{L} - \gamma\sub{L}^2)^2}. \label{eq:260}
\end{equation}
The right-hand side takes a maximum value for $|\varDelta| =
\sqrt{2\gamma_0\gamma\sub{L}+\gamma\sub{L}^2}$, which we define as
$\varDelta\sub{max}$. 
In the case of $\gamma\sub{L} = 0$, the
transmission bandwidth can become infinitesimal and the group index can
become infinite. Note that the transmission bandwidth can be smaller
than the resonance linewidth $\gamma_0$ of the meta-atoms when the leak
$\gamma\sub{L}$ of the indirect coupling is sufficiently small. 
It is also found from Eq.\,(\ref{eq:260}) that the group index becomes
negative; that is, the transmission window disappears in the range
$|\varDelta| < \gamma\sub{L}$.

Here we discuss the derivation of the leak $\gamma\sub{L}$ of the
indirect coupling, which is an important parameter that determines the
minimum bandwidth of the transparency window. 
Since the indirect coupling is mediated by radiative modes, the indirect
coupling takes a maximum value, that is, $\gamma\sub{L}$ vanishes when
all the energy dissipated from one meta-atom in each unit cell is
absorbed by the other meta-atom. 
The dissipated energy is derived from Ohmic
loss, dielectric loss, and radiative loss in most meta-atoms. 
The dissipated energy derived
from Ohmic and dielectric losses
cannot excite the other meta-atom, and thus, Ohmic and dielectric losses
contribute
to $\gamma\sub{L}$. In addition, 
the difference between the radiation modes of the
two kinds of meta-atoms causes a reduction of the radiative coupling,
i.e., an increase in $\gamma\sub{L}$. 
Thus, $\gamma\sub{L}$ is derived from Ohmic loss, dielectric loss, 
and the difference
between the radiation modes of the two kinds of meta-atoms.

We also discuss the physical meaning of the narrower transmission bandwidth
than the resonance linewidths of the meta-atoms 
in the transparency phenomenon.
For $\omega = \omega_0$, Eq.\,(\ref{eq:40}) is reduced to 
$x_1 \approx -x_2$. In this case, 
the dissipation terms derived from $\gamma_{1,2}$ 
in Eqs.\,(\ref{eq:10}) and (\ref{eq:20}) are (partially) 
canceled out by the terms
derived from the indirect coupling $\kappa^2 = \ii \Im{(\kappa^2)}$. 
This implies that the energy radiated from one meta-atom in each unit
cell is absorbed by
the other meta-atom. That is, the radiated energy moves backward and
forward between the two kinds of meta-atoms in each unit cell. 
Therefore, the effective radiation loss in the metamolecule is reduced and
the narrowband transparency phenomenon can be achieved. 

We assumed $\Re{(\kappa^2)}=0$ in the above discussion. In the case of
$\Re{(\kappa^2)} \neq 0$, it is found from Eq.\,(\ref{eq:30}) that the
EIT-like transparency phenomenon occurs for $\omega^2 = \omega_0^2 +
\Re{(\kappa^2)}$. 

It is useful for understanding the electromagnetic response of the
metamaterial to analyze an electrical circuit model of the
metamolecule. We consider an electrical circuit that consists of two
coupled inductor--capacitor resonant circuits, shown in
Fig.\,\ref{fig:model}(b). Three kinds of couplings exist in the
electrical circuit. Applying Kirchhoff's voltage law 
for the electrical circuit yields the
following equations: 
\begin{align}
& \left( - \omega^2 + \frac{1}{L C_1} - \ii \frac{R_1}{L} \omega
 \right) q_1 \nn \\
& \hspace{5mm} - \frac{\ii \omega}{L} 
\left[ R\sub{M} -\ii \left( \omega M - \frac{1}{\omega C\sub{M}} 
\right) \right]
q_2 = \frac{V_1}{L},  \label{eq:60} \\
& \left( - \omega^2 + \frac{1}{L C_2} - \ii \frac{R_2}{L} \omega
 \right) q_2 \nn \\
& \hspace{5mm} - \frac{\ii \omega}{L} 
\left[ R\sub{M} -\ii \left( \omega M - \frac{1}{\omega C\sub{M}} 
\right) \right]
q_1 = \frac{V_2}{L}, \label{eq:70}
\end{align}
where $L_1 = L_2 = L$ is assumed. 
It is found by comparing Eqs.\,(\ref{eq:10}) and (\ref{eq:20}) with
Eqs.\,(\ref{eq:60}) and (\ref{eq:70}) that the imaginary and real parts of 
$\kappa^2$ correspond to the real and imaginary parts, respectively, of the mutual impedance 
$Z\sub{M} = R\sub{M} - \ii [ \omega M - (\omega C\sub{M})^{-1}]$. 
This relation shows 
the influences of the electric coupling $C\sub{M}$, magnetic coupling
$M$, and energy coupling $R\sub{M}$ on $\kappa^2$. 

We now discuss the difference between the EIT-like metamaterial with 
indirect coupling and that with direct
coupling.\cite{zhang_prl08,tassin_prl09,liu_nat09,tamayama10,zhang_apl_10,kurter11,tamayama12}  
The unit cell of the latter metamaterial consists of two
directly coupled meta-atoms: a low quality-factor meta-atom,
which interacts with the incident electromagnetic wave, and
a high quality-factor meta-atom, which does not interact with the incident
wave.
The loss in the high quality-factor meta-atom should be reduced to achieve a
large group index. The resonant frequencies of the two kinds of 
meta-atoms should
be identical to ensure that the transmission spectrum is symmetric
to reduce higher-order dispersion. 
On the other hand, as described above, the unit cell of 
the former metamaterial consists of two similar meta-atoms.
The indirect coupling should be strong to realize a large group
index. The two meta-atoms should be coupled so that $\Re{(\kappa^2)}=0$
is satisfied to ensure that the transmission spectrum is symmetric. 
Both kinds of EIT-like metamaterials require efforts to
achieve a large group index and symmetric transmission spectrum. 
However, the structure of the meta-atoms can be simple for
the former metamaterial, because the unit cell consists of two similar
meta-atoms and the radiation losses of the meta-atoms may be large. 
Therefore, the former metamaterial can be superior in terms of ease of
design and fabrication to the latter
metamaterial
if $\gamma\sub{L} \simeq 0$ and $\Re{(\kappa^2)}=0$ are
simultaneously satisfied.

\section{FDTD analysis of EIT-like metamaterial with indirect coupling}

\begin{figure}[tb]
\begin{center}
\includegraphics[scale=1]{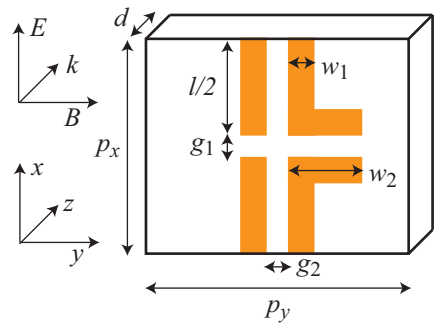}
\caption{Unit cell of the EIT-like metamaterial with indirect coupling. 
A thin metallic film represented by brown (gray)
lies on a dielectric substrate represented by white. }
\label{fig:unit}
\end{center}
\end{figure}

\begin{figure}[tb]
\begin{center}
\includegraphics[scale=1]{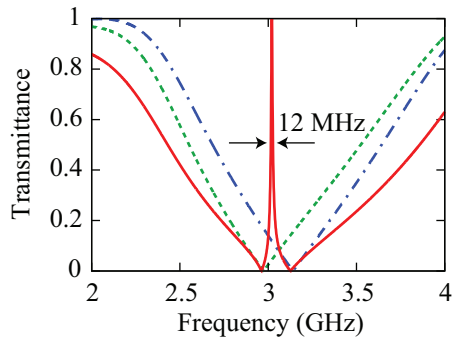}
\caption{Transmission spectra for the lossless substrate. The
 red solid curve represents the transmission spectrum of the 
 metamaterial composed of coupled cut-wire pairs. 
 The green dashed curve and blue dashed-dotted curve
 represent the transmission spectra of the metamaterials composed of one
 kind of cut wires.}
\label{fig:trans_lossless}
\end{center}
\end{figure}

In order to confirm the validity of the above theory and to investigate
whether $\gamma\sub{L} \simeq 0$ and $\Re{(\kappa^2)}=0$ are
simultaneously satisfied, we design the EIT-like metamaterial with
indirect coupling and analyze the characteristics of the 
metamaterial using an FDTD method.\cite{taflove05} 
Figure \ref{fig:unit} shows the unit cell of the EIT-like metamaterial
whose electromagnetic 
response is modeled by the mechanical model shown in
Fig.\,\ref{fig:model}(a). Two kinds of cut-wire resonators 
(meta-atoms) with different
resonant frequencies are placed with a gap
of $g_2$. The structures of the two kinds of cut-wire resonators are
designed to be similar to each other in order to make their characteristics
including the radiation modes similar. 
The configuration of the two kinds of 
cut-wire resonators is determined so that 
the radiation from one cut-wire resonator can excite the other cut-wire
resonator; that is, indirect coupling is induced in the cut-wire
pair. 
The unit structure can be regarded as a kind of asymmetric
split-ring resonator.\cite{fedotov07,plum09} 
The resonant frequencies, $\omega_1$ and $\omega_2$, of the two kinds of
resonators are determined by the inductance
derived from the metal wire with length $l$ and the capacitance
derived from the gap $g_1$. The difference $| \varDelta |$ of the resonant
angular frequencies between the two kinds of resonators 
is determined by the difference
between $w_1$ and $w_2$. The resonance linewidths, $\gamma_1$ and
$\gamma_2$, of the 
two kinds of resonators 
depend on the radiation loss, dielectric loss in the substrate, and 
Ohmic loss in the metal. 
The coupling factor $\kappa$ between the two kinds of resonators can be
controlled by varying $g_2$. 
The FDTD simulation is performed in the microwave region where metals
can be regarded 
as perfect electric conductors; that is, Ohmic loss in the metal is
negligible. 
This enables us to investigate the influence of the leak $\gamma\sub{L}$
of the
indirect coupling on the transmission characteristics 
by only varying the dielectric loss of the substrate and to easily 
compare the
theory based on the mechanical model with the results of the FDTD
analysis. 

\begin{figure}[tb]
\begin{center}
\includegraphics[scale=0.8]{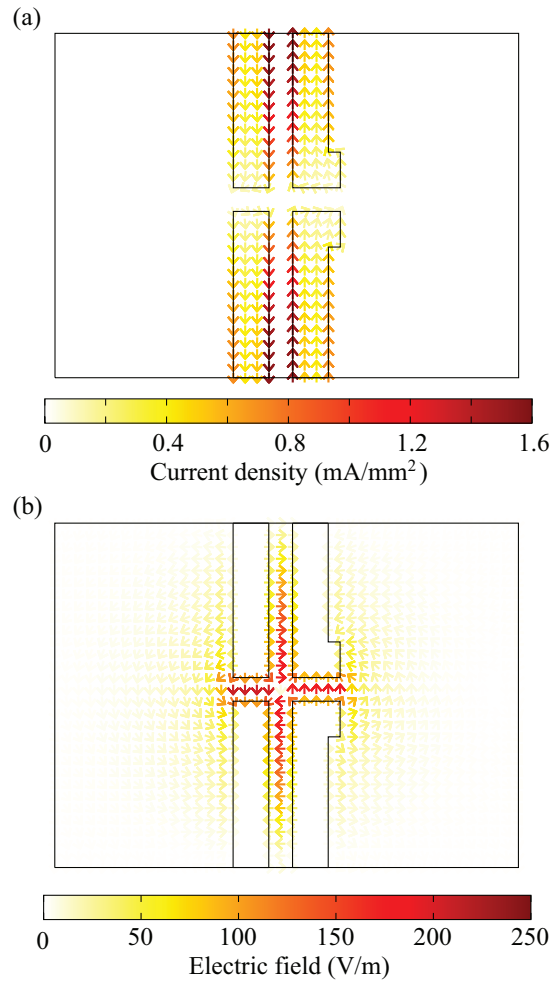}
\caption{Snapshots of (a) the current and (b) electric field
 distributions at the transparency frequency. The incident electric
 field is $1\,\U{V/m}$.}
\label{fig:field}
\end{center}
\end{figure}

We first calculated the transmission spectrum of the EIT-like
metamaterial. The geometrical parameters of the metamaterial were set to
$l = 28\,\U{mm}$, $w_1 = 3\,\U{mm}$,
$w_2 = 4\,\U{mm}$, $g_1 = 2\,\U{mm}$, 
$g_2 = 2\,\U{mm}$, $d =1 \,\U{mm}$, $p_x = 29\,\U{mm}$, and
$p_y = 39\,\U{mm}$. The relative permittivity of the substrate was 
set to 3.3, which is the real part of the relative permittivity
of polyphenylene ether. 
The FDTD simulation space was discretized into uniform cubes with
dimensions of $1\,\U{mm} \times 1\,\U{mm} \times 1\,\U{mm}$. 
(Although the thickness of the substrate was modeled using only a single
FDTD cell in the simulation, no significant errors were caused to the
simulation results.) 
Periodic boundary conditions were
applied to the $x$ and $y$ directions to realize periodically arranged
metamolecules. 

Figure \ref{fig:trans_lossless} shows the
transmission spectra of three kinds of metamaterials. 
The red solid curve represents the transmission spectrum of the
metamaterial composed of coupled cut-wire pairs (metamolecules)
described above. The green dashed curve (blue dashed-dotted curve)
represents the transmission spectrum of the metamaterial composed of one
kind of cut wires, i.e., meta-atoms, shown in the 
right-hand side (left-hand
side) of Fig.\,\ref{fig:unit}. While simple
absorption spectra are observed for the metamaterials composed of one
kind of cut wires, a transmission window is observed in the 
absorption spectrum for the metamaterial composed of coupled cut-wire
pairs. In addition, the transmission bandwidth is much smaller than the
resonance linewidths (absorption bandwidths) of the metamaterials composed
of one kind of cut wires. This implies that the EIT-like transparency
phenomenon occurs for the metamaterial composed of coupled cut-wire
pairs owing to the indirect coupling. 
Since the transmittance at the transparency frequency is nearly unity, 
$\gamma\sub{L}$ seems to be negligibly small for the
present geometrical parameters. 
Although the present metamaterial, which is 
composed of periodically arranged 
coupled resonators, is different from  
isolated coupled resonators such as the mechanical model shown in
Fig.\,\ref{fig:model}(a) and the
coupled resonators in 
previous studies,\cite{suh04,verslegers12,zhang_s_12} indirect
coupling can be induced between the meta-atoms in each unit cell and 
$\gamma\sub{L}$ seems to be negligibly small also in the 
metamaterial. (Note that it is still unclear whether indirect coupling
is induced between different cells.)

We next calculated the field distributions at the transparency frequency
to understand the physical meaning of the EIT-like transparency
phenomenon. Figure \ref{fig:field}(a) shows the current distribution at
the transparency frequency. An antisymmetric current flows in the
coupled cut-wire pair and thus the total electric dipole moment vanishes. 
That is, the scattering is suppressed due to
destructive interference between the radiations from the two kinds of
cut wires. This observation is another aspect of the cancellation
between the radiation loss and the indirect coupling described in
Sec.\,II. 
Figure \ref{fig:field}(b) shows the electric field
distribution at the transparency frequency. A large quadrupole electric
field is induced at the gap of the two kinds of cut wire. 
The electric field at the gap is about 200 times as large as the
incident electric field, in which the narrow band effect is reflected. 

\begin{figure}[tb]
\begin{center}
\includegraphics[scale=1]{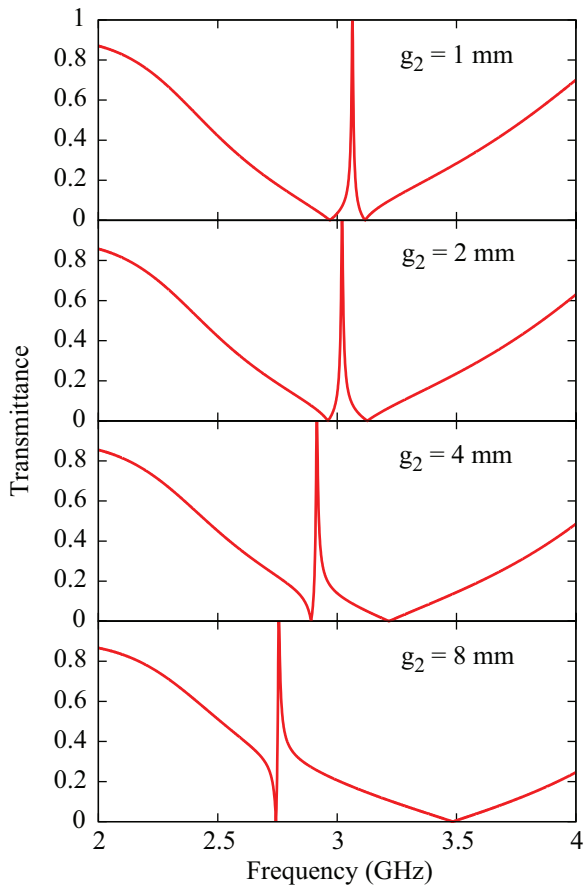}
\caption{Dependence of the transmission spectrum on $g_2$. }
\label{fig:trans_g12}
\end{center}
\end{figure}

We next analyzed the dependence of the transmission spectrum on $g_2$ to
investigate whether $\Re{(\kappa^2)}=0$ can be satisfied by varying $g_2$. 
Figure \ref{fig:trans_g12} shows the transmission spectra for $g_2 =
1\,\U{mm}$, 2\,mm, 4\,mm, and 8\,mm. The other parameters are the same
as those in the case of Fig.\,\ref{fig:trans_lossless}. 
With increasing $g_2$, the transmission peak shifts to lower frequency; 
that is, $\Re{(\kappa^2)}$ decreases. 
The transmission spectrum for $g_2=1\,\U{mm}$ shows the opposite asymmetry
to that for $g_2=2\,\U{mm}$ and, therefore, the condition that satisfies
$\Re{(\kappa^2)}=0$ exists in the range $1\,\U{mm} < g_2 <
2\,\U{mm}$. The imaginary part of $\kappa^2$ seems to be a constant
value in the range $1\,\U{mm} < g_2 < 8\,\U{mm}$ because the
transmittance at the transparency frequency is nearly unity for these four
calculated conditions. 

The dependence of $\Re{(\kappa^2)}$ on $g_2$ can be understood using the
electrical circuit model of the coupled resonators 
shown in Fig.\,\ref{fig:model}(b). 
As $g_2$ increases, 
both the mutual inductance $M$ and mutual capacitance $C\sub{M}$
decrease and thus the imaginary part of the mutual impedance $Z\sub{M}$
decreases. Therefore, $\Re{(\kappa^2)}$ decreases and the transmission
peak shifts to lower frequency with increasing $g_2$. 
Note that this discussion can be applied only to the case of $g_2 \ll
\lambda$, where $\lambda$ is the wavelength of the electromagnetic
waves.  
When $g_2$ is comparable to or larger than $\lambda$, the phase
retardation of the coupling has to be taken into account;
therefore, we cannot discuss the indirect and direct couplings
separately. However, $g_2 \ll \lambda$ is safely satisfied in
metamaterials.

\begin{figure}[tb]
\begin{center}
\includegraphics[scale=1]{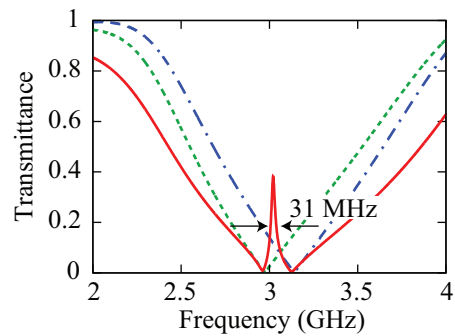}
\caption{Transmission spectra for $\tan\delta = 0.005$. The
 red solid curve represents the transmission spectrum of the 
 metamaterial composed of coupled cut-wire pairs. 
 The green dashed curve and blue dashed-dotted curve
 represent the transmission spectra of the metamaterials composed of one
 kind of cut wires.}
\label{fig:trans_lossy}
\end{center}
\end{figure}

\begin{figure}
\begin{center}
\includegraphics[scale=1]{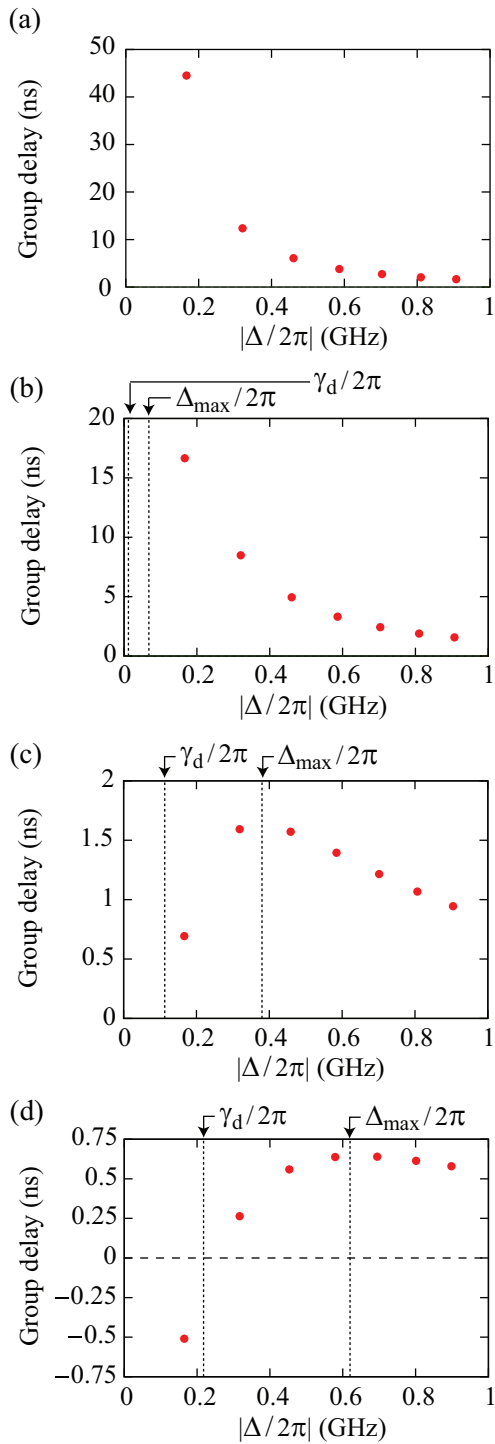}
\caption{Dependence of the group delay at the transparency frequency 
 on $| \varDelta |$ when $\tan\delta$ is equal to 
 (a) 0, (b) 0.005, (c) 0.05, and (d)
 0.1. The values of $\varDelta\sub{max}$ and $\gamma\sub{d}$ are
 represented by the vertical dashed lines in each case. Note that the
 scales of the vertical axes are different from each other.}
\label{fig:group_delay}
\end{center}
\end{figure}

We next analyzed the influence of the dielectric loss of the substrate on the
transmission characteristics. Figure
\ref{fig:trans_lossy} shows the transmission spectra of the 
above mentioned three kinds of metamaterials when the loss
tangent $\tan\delta$ 
of the dielectric substrate is 0.005, which is the loss tangent
of polyphenylene ether at 3\,GHz. The other parameters are
the same as those in the case of Fig.\,\ref{fig:trans_lossless}. 
While the transmission
spectra of the metamaterials composed of one kind of cut wires are
almost the same as those in Fig.\,\ref{fig:trans_lossless}, 
the transmittance of
the metamaterial composed of coupled cut-wire pairs at
the transparency frequency is smaller than that in the case of
Fig.\,\ref{fig:trans_lossless} due to the dielectric loss. 
This implies that the
leak of the indirect coupling without dielectric loss 
is much smaller than the dielectric loss
$\gamma\sub{d}$ of the substrate. Thus, we may assume $\gamma\sub{L}
\simeq \gamma\sub{d}$ when $\tan\delta$ is larger than
0.005 at most. 

Finally, 
we calculated the dependence of the group delay at the
transparency frequency on $| \varDelta |$ to investigate the influence
of $\tan\delta$ on $\varDelta\sub{max}$. 
The absolute value of 
$\varDelta$ was varied by varying $w_2$ from $4\,\U{mm}$ to $10\,\U{mm}$
with steps of $1\,\U{mm}$. 
The other parameters except $\tan\delta$
are the same as those in the case of
Fig.\,\ref{fig:trans_lossless}. 

Figure \ref{fig:group_delay} shows the group delay at the transparency
frequency as a function of
$| \varDelta |$ for $\tan\delta =  0$, 0.005, 0.05, and 0.1. 
For the calculated conditions, 
the group delay monotonically increases with decreasing $| \varDelta |$ 
for $\tan\delta = 0$ and 0.005, while the group
delay first increases and then decreases with decreasing $| \varDelta |$
for $\tan\delta = 0.05$ and 0.1.

We now discuss the above results using the mechanical model of the coupled
resonator shown in Fig.\,\ref{fig:model}(a). We have to
evaluate in advance the radiation loss $\gamma_0$ of the 
cut-wire resonators and the dielectric loss $\gamma\sub{d}$ of the
substrate in order to use the mechanical model. 
For a Lorentz medium with a simple absorption line, $\gamma_0$ is 
almost equal to the bandwidth of the negative group delay. Thus,
$\gamma_0$ can 
be estimated by calculating the group delay of the metamaterial composed
of one kind of cut wires (not shown). The value of 
$\gamma\sub{d}$ can be estimated as follows. 
In a series inductor--capacitor resonant circuit, the quality factor of the
circuit can be approximated as the inverse of the loss tangent of the
dielectric in the capacitor when the loss in the
circuit is caused only by the dielectric loss in the capacitor. 
Since the
thin metallic film is on the dielectric substrate in the metamaterial
shown in Fig.\,\ref{fig:unit}, we assume that
half of the capacitor in the metamaterial is filled with the
dielectric and the other half is filled with vacuum. 
That is, the capacitance of the gap is assumed to be 
$(1+\varepsilon\sub{r})/2$ times as
large as that without the substrate, where $\varepsilon\sub{r}$ is the
relative permittivity of the substrate. 
From this assumption, 
the effective loss tangent
$\tan\delta^{\prime}$, which is the loss tangent when the capacitor is
assumed to be filled with a uniform medium, 
is found to be 
$\{ \Re{(\varepsilon\sub{r})} / [ 1 + \Re{(\varepsilon\sub{r})}] \} 
\tan\delta$. 
Therefore, the dielectric loss $\gamma\sub{d}$ in the metamaterial is 
estimated to be $\omega_0 \tan\delta^{\prime}$. 

We now show $\varDelta\sub{max}$ calculated using the above
estimated $\gamma_0$ and
$\gamma\sub{d}$ in Fig.\,\ref{fig:group_delay}. 
The value of $\gamma\sub{d}$ is also shown in the figure. 
It is found that the group delay increases or decreases with decreasing
$| \varDelta |$ when $| \varDelta |$ is larger or smaller, respectively, than
$\varDelta\sub{max}$ and that the group delay seems to vanish at around
$|\varDelta| = \gamma\sub{d}$. This
implies that the behavior of the coupled cut-wire pair metamaterial can be
well described by the mechanical model. 
That is, indirect coupling can be induced only between meta-atoms in
each unit cell of metamaterials composed of
periodically arranged coupled resonators.

\section{Conclusion}

We analyzed the EIT-like transparency phenomenon in metamaterials
composed of coupled resonators with indirect coupling. The theoretical
analysis based on the mechanical model showed that the transparency
bandwidth can be narrower than the resonance linewidths of the
constitutive resonators when strong indirect coupling is
introduced. The FDTD simulation demonstrated that the EIT-like
transparency
phenomenon with $\gamma\sub{L} \simeq 0$ and $\Re{(\kappa^2)}=0$ 
can occur in metamaterials composed of coupled
cut-wire pairs. 
The characteristics of the metamaterial was
confirmed to be well described by the mechanical model, and indirect
coupling was found to be induced only between meta-atoms in each unit
cell of metamaterials composed of
periodically arranged coupled resonators. 

Structures of meta-atoms with narrow resonance linewidth usually have
complicated shapes. However, the narrow band
transparency window can be obtained using simple structures such as
cut wires when indirect coupling is introduced. If dielectric
loss is prevented completely, an
extremely narrow transparency window can be achieved
below infrared frequencies where metals can be
regarded as perfect electric conductors. 
For future studies, the minimum transparency bandwidth of the
EIT-like metamaterial without dielectric loss 
needs to be investigated experimentally. 
In the optical region, metals exhibit relatively large Ohmic losses, and thus
the metamaterial should be designed with low-loss dielectrics. 
Indirect coupling would be useful not only for realizing
EIT-like phenomena but also for other techniques for controlling
electromagnetic waves. 

\begin{acknowledgments}

This research was supported by a Grant-in-Aid for Scientific Research on
Innovative Areas (No.\@ 22109004) from the Ministry of
Education, Culture, Sports, Science, and Technology, Japan, and by a
Grant-in-Aid for Research Activity Start-up (No.\@ 25889028) from the Japan
Society for the Promotion of Science. 

\end{acknowledgments}

%\bibliography{main}

%merlin.mbs apsrev4-1.bst 2010-07-25 4.21a (PWD, AO, DPC) hacked
%Control: key (0)
%Control: author (8) initials jnrlst
%Control: editor formatted (1) identically to author
%Control: production of article title (-1) disabled
%Control: page (0) single
%Control: year (1) truncated
%Control: production of eprint (0) enabled
%

\end{document}